\documentclass[aps,twocolumn,pra,tightenlines,floatfix,showpacs]{revtex4}

\usepackage[dvips]{graphicx}
\usepackage[english]{babel}
\usepackage{amsmath}
\usepackage{amssymb}
\usepackage{times}

\newcommand{\bs}{\begin{split}}
\newcommand{\es}{\end{split}}
\newcommand{\be}{\begin{equation}}
\newcommand{\ee}{\end{equation}}
\newcommand{\ba}{\begin{eqnarray}}
\newcommand{\ea}{\end{eqnarray}}

\begin{document}

\title{Ground state description of a single vortex in an atomic Fermi
  gas: From BCS to Bose-Einstein condensation}

\author{Chih-Chun Chien, Yan He, Qijin Chen, and K.  Levin}

\affiliation{James Franck Institute and Department of Physics,
 University of Chicago, Chicago, Illinois 60637}

\date{\today}

\begin{abstract}
  We use a Bogoliubov-de Gennes (BdG) formulation to describe a single
  vortex in a neutral fermionic gas.  It is presumed that the attractive
  pairing interaction can be arbitrarily tuned to exhibit a crossover
  from BCS to Bose-Einstein condensation.  Our starting point is the
  BCS-Leggett mean field ground state for which a BdG approach is
  microscopically justified.  At strong coupling, we demonstrate that
  this approach is analytically equivalent to the Gross-Pitaevskii
  description of vortices in true bosonic systems.  We analyze the
  sizable density depletion found for the unitary regime
  and relate it to the presence of unoccupied (positive energy)
  quasi-bound states at the core center.
\end{abstract}

\pacs{03.75.Hh, 03.75.Ss, 74.20.-z \hfill \textsf{\textbf{cond-mat/0510647}} }

\maketitle

One of the most exciting developments in atomic and condensed matter
physics has been the observation of superfluidity in trapped fermionic
systems \cite{Jin4,Ketterle3a,Thomas2a,Grimm3a}.  In these systems, the
presence of a Feshbach resonance provides a means of tuning the
attractive pairing interaction with applied magnetic field.  In this way
the system undergoes a continuous evolution from BCS to Bose-Einstein
condensed (BEC) superfluidity.

The most conclusive demonstration of the superfluid phase has been the
experimental observation of vortices \cite{KetterleV}.  Particularly
interesting from a theoretical viewpoint is the way vortices evolve from
BCS to BEC. This evolution is associated, not just with a decrease in
vortex size but with a complete rearrangement of the fermionic states
which make up the core. As a result, there is a continuous evolution of
the particle density within a vortex, thereby affecting the visibility
of vortices in the laboratory.  In this paper we discuss the behavior of
a (single) vortex as the system crosses from BCS to BEC.  Our work is
based on simplest BCS-like ground state first introduced by Leggett
\cite{Leggett} and Eagles \cite{Eagles} to treat BCS-BEC crossover.
With this choice of ground state inhomogeneity effects are readily
incorporated as in generalized Bogoliubov-de Gennes (BdG) theory.  Here
we demonstrate analytically that the BdG strong coupling description of
the $T=0$ vortex state coincides with the usual Gross-Pitaevskii (GP)
treatment of vortices in bosonic superfluids.  A fermionic theory based
on BdG is, thus, very inclusive, and within this approach one expects a
smooth evolution of vortices from the BCS  to BEC limit as the statistics
effectively change from fermionic to bosonic.

Previous studies of vortices in these fermionic superfluids addressed
the BCS limit at $T=0$ \cite{NB03} and $ T \approx T_c$ \cite{RPT01}.
There is also work \cite{BY03} on the $T=0$ strict unitary case where a
BdG approach was used with Hartree-Fock contributions included.  In the
present work, by contrast, we discuss the entire crossover regime and,
importantly, present a detailed analysis of the energy and spatial
structure within the core and how it evolves from BCS to BEC.
A very different path integral approach was introduced in
Ref.~\cite{Devreese} to address vortices with BCS-BEC crossover, but
here the authors note that density depletion effects appear to be
unphysically large in the BCS regime.  Our analytical approach builds
heavily on previous work \cite{PSP03} which showed a general connection
between GP theory and BdG. From this one can conclude that a generalized
BCS theory \cite{Leggett} treats the bosonic degrees of freedom at the
same level as GP theory.  Different ground states can be
contemplated, (with incomplete condensation, say) but they will not be
compatible with BdG theory.  In a similar way, once $T \neq 0$ one has
to incorporate noncondensed pairs, and associated pseudogap physics
\cite{CSTL05} which are not present in a finite temperature BdG theory.

For the most part, BdG approaches require detailed numerical solution
\cite{BK69,GS91,NB03,Griffin5}, so it is particularly useful to have
analytical tools in the BEC limit.  We present this non-numerical
description first.  Our general self consistent equations \cite{DG66}
are
\begin{equation}\label{eq:BdG}
\left( \begin{array}{cc} h-\mu & \Delta (\mathbf{r}) \\ 
\Delta^{*}(\mathbf{r}) & -h^{*}+\mu \end{array} \right) 
\left( \begin{array}{c} u_{n} \\ v_{n} \end{array}\right) 
= E_{n}\left(\begin{array}{c} u_{n} \\ v_{n} 
\end{array}\right), 
\end{equation}
where $h=-\frac{1}{2m}\nabla^{2}+V_{ext}(\mathbf{r})$,
$\Delta(\mathbf{r})$ is the $T=0$ gap function which is importantly the
same as the $T=0$ order parameter, $\Delta_{sc}(r)$,
$V_{ext}(\mathbf{r})$ is the external potential associated with the
trap, and we choose $\hbar=1$ with $\int
d\mathbf{r}(|u_{n}|^{2}+|v_{n}|^{2})=1$ for all energy levels $n$.  The
difference between the present approach and the usual BdG applications
of superconductivity is that here the fermionic chemical potential $\mu$
must be self consistently determined, as the attractive coupling
constant is varied.

We use a Green's function formulation \cite{BK69} to write the
zero-temperature 
free energy $E_{0}=\langle H-\mu N\rangle$,
(where $N$ is the number operator) in the form
\begin{eqnarray}\label{eq:GsEG}
E_{0}&=&\int d\mathbf{r} \Big(\frac{|\Delta(\mathbf{r})|^{2}}{V} \nonumber \\
& &{}+\int_{-\infty}^{\infty}\frac{d\omega}{2\pi} \int_{\omega}^{\infty
  sgn(\omega)}d\omega^{\prime} Tr[i\tau_{3}\hat{G}
(\mathbf{r},\mathbf{r},\omega^{\prime})] \Big), 
\end{eqnarray}
Here $\tau_{i}$ (i=1,2,3) are the  Pauli matrices.
The elements of $\hat{G}$ corresponding to the normal
and anomalous channels can be further expressed as 
coupled integral equations in terms of the noninteracting 
Green's function $G_{11}^{(0)}$ 
\begin{eqnarray}\label{eq:G11}
G_{11}(\mathbf{r},\mathbf{r^{\prime}};\omega) &=&
G_{11}^{(0)}(\mathbf{r},\mathbf{r^{\prime}};\omega) +\int
d\mathbf{r}_{2}[G_{11}^{(0)}(\mathbf{r},\mathbf{r}_{2};\omega) \nonumber
\\ 
&&{}\times \Delta(\mathbf{r}_{2})
G_{21}(\mathbf{r}_{2},\mathbf{r^{\prime}};\omega)] \,, 
\end{eqnarray}
%
\begin{equation}\label{eq:G21}
G_{21}(\mathbf{r},\mathbf{r^{\prime}};\omega) = -\int
d\mathbf{r}_{2}G_{11}^{(0)}(\mathbf{r}_{2},\mathbf{r};-\omega)
\Delta^{*}(\mathbf{r}_{2})G_{11}(\mathbf{r}_{2},\mathbf{r^{\prime}};\omega).  
\end{equation}
where the gap and number equation 
are given by $\Delta^{*}(\mathbf{r})=-V\int_{-\infty}^{\infty}\frac{d\omega}
{2\pi}G_{21}(\mathbf{r},\mathbf{r};\omega)$ and $n(\mathbf{r})=
2\int_{-\infty}^{\infty}\frac{d\omega}{2\pi}
e^{i\omega\eta}G_{11}(\mathbf{r},\mathbf{r};\omega)$. 
The coupling constant $-V<0$ for the attractive inter-fermion contact
interaction is parametrized in terms of the $s$-wave scattering length
$a_{F}$ with $m/4\pi a_{F}=-1/V+\sum_{k}m/k^{2}$.

In the strong pairing limit $a_{F}$ is small and positive and
$\Delta/|\mu|$ is small.  To derive the ground state energy $E_{0}$
from Eq.(\ref{eq:GsEG}), we expand the Green's function $\hat{G}$ in the
gap equation in terms \cite{PSP03} of $G_{11}^{(0)}$.
Including terms up to fourth order in $|\Delta|$, it
follows that
\begin{eqnarray}
  E_{0}[\Delta]&=&\int d\mathbf{r}\Big\{
  \frac{|\Delta(\mathbf{r})|^{2}}{V} -
  a_{0}(\mathbf{r})|\Delta(\mathbf{r})|^{2} \nonumber\\
  &&{}+\frac{1}{2}b_{0}(\mathbf{r})| \nabla
  \Delta(\mathbf{r})|^{2}-\frac{1}{2}c_{0}(\mathbf{r})|
  \Delta(\mathbf{r})|^{4} \Big\}\,
\end{eqnarray}
where $a_{0}(\mathbf{r})\simeq\frac{1}{V}+
\frac{m^{2}a_{F}}{8\pi}[\mu_{B}-2V_{ext}(\mathbf{r})]$,
$b_{0}(\mathbf{r})\simeq ma_{F}/16\pi$, and
$c_{0}(\mathbf{r})\simeq - m^{3}a_{F}^{3}/16\pi$.  Here
$\mu_{B}=2\mu+\varepsilon_{0}$ is the effective ``bare'' bosonic
chemical potential, and $\varepsilon_{0}=(ma_{F}^{2})^{-1}$ is the
binding energy of the composite boson, with $|\mu_{B}|\ll\varepsilon_{0}$.
It is assumed that $V_{ext}$ is slowly varying \cite{PSP03} so that
$G_{11}^{(0)}(\mathbf{r},\mathbf{r^{\prime}};\omega_{s})\ \simeq
G_{11}^{(0)}(\mathbf{r}-\mathbf{r^{\prime}}; \mu- [V_{ext}(\mathbf{r})+V_{ext}
(\mathbf{r^{\prime}})]/2)$.  Similarly, $\Delta(\mathbf{r})$ is assumed
to vary slowly on the scale of $a_{F}$.  As a consequence of these
assumptions, and for the purposes of this paper, (which ultimately
focuses on a single vortex), trap effects are not particularly relevant.

It should be stressed that this expansion is similar to Gor'kov's
derivation of Ginzburg-Landau (GL) theory in the BCS limit at $T \approx
T_c$, albeit here we consider strong coupling and $T=0$.  As in
conventional superconductors \cite{P69}, minimizing the energy
$E_{0}[\Psi]$ with respect to $\Psi$, one obtains
\begin{equation}
  E_{0}[\Psi]= -\frac{2\pi a_{B}}{m_{B}}\int d\mathbf{r}\,\, |\Psi|^{4} 
+\frac{1}{2m_{B}}  \int \mathrm{d}\mathbf{l}\,\, 
\Psi^{*}\hat{n}\cdot\nabla \Psi|_{S} .
\end{equation}
where we have identified the condensate wave function
$\Psi(\mathbf{r})$ as $\sqrt{m^{2}a_{F}/8\pi}\,\,\Delta(\mathbf{r})$
and $\hat{n}$ is the unit vector normal to the surface.  The second
term is a surface term, which vanishes for an infinite system. In a
neutral superfluid, however, the energy has to be calculated within a
finite region of radius $R$ around the vortex core, to avoid
divergences, so this surface term cannot be neglected.

Equivalently, the zero-temperature energy can be written in a more
conventional form as 
\begin{eqnarray}
\label{eq:E0}
E_{0}[\Psi]&=&\int d\mathbf{r}\, \Big\{\frac{1}{2m_{B}}| 
\nabla\Psi(\mathbf{r})|^{2} + 2V_{ext}(\mathbf{r})| 
\Psi(\mathbf{r})|^{2} \nonumber \\
& &{}+\frac{1}{2}U_{0}|\Psi(\mathbf{r})|^{4} - 
\mu_{B}|\Psi(\mathbf{r})|^{2}\Big\}, 
\end{eqnarray}
where $U_{0}=4\pi a_{B}/m_{B}$, $m_{B}=2m$ and $a_{B}=2a_{F}$ are the
mass and the scattering length of the composite boson.

\begin{figure*}[t]
\centerline{\includegraphics[clip,width=5.in]{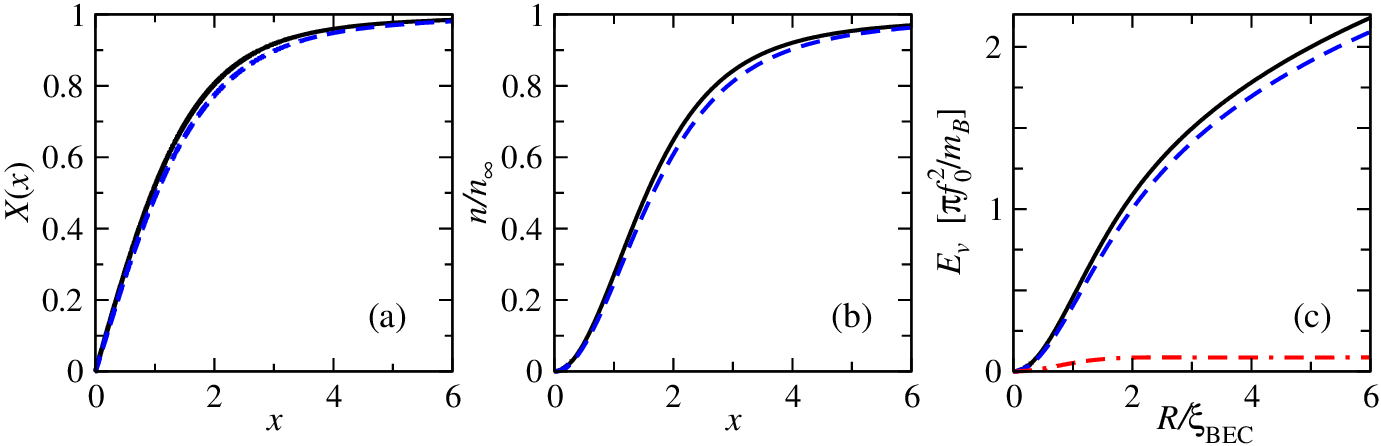}}
\caption{(a) Numerical solution of GP equation in the BEC limit with
  ($g_{3}^{\prime}=0.1$, dashed lines) and without (solid lines) the
  three-body $g_{3}$-term, (b) the corresponding normalized particle
  density $n(x)/n_\infty$ as a function of $x$, and (c) the
  zero-temperature vortex energy cost $E_{v}$ as a function of
  $R/\xi_{BEC}$. In (c), the difference is shown as the red dot-dashed
  curve . Here $n_\infty \equiv n(\infty)$.}
\label{fig:XEvwg3}
\end{figure*}

Importantly, this expression has the same form as the $T=0$ energy of a
gas of weakly-interacting bosons \cite{PSB03} associated with GP theory.
It should be noted that here this GP theory is written in terms of the
grand canonical representation where the bosonic chemical potential
(rather than the number of particles $N$) is held fixed.  Minimizing the
zero-temperature energy $E_{0}$ (via $\delta E_{0}/\delta\Psi^{*}=0$),
leads to the well known GP equation:
\begin{equation}\label{eq:GP}
-\frac{1}{2m_{B}}\nabla^{2}\Psi(\mathbf{r})+2V_{ext}(\mathbf{r}) +
U_{0}|\Psi(\mathbf{r})|^{2}\Psi(\mathbf{r})=\mu_{B}\Psi(\mathbf{r}). 
\end{equation}

We emphasize that this BdG analysis has, in effect, derived GP theory
from a fermionic starting point.  The presence of a Hartree term in the
BdG equations will destroy the simple analytic arguments presented here.
Nevertheless, a Hartree contribution for the composite bosons is found
here of the form $U_{0}|\Psi|^{2}\Psi \rightarrow U_{0}n_{B}\Psi$,
(where $n_{B}$ is the density of bosons). This is, of course, unrelated
to the Hartree term of the original fermions, which is absent in
Eq.(\ref{eq:BdG}), as is consistent with the usual BCS-Leggett mean
field theory \cite{Leggett}.  The inclusion of Hartree terms \cite{NB03}
in the vortex problem has been accounted for in the literature, at weak
\cite{NB03} and strict unitary coupling \cite{BY03}, but they do not
appear to lead to important differences.

Expanding the zero-temperature energy $E_{0}$ to next order in $\Delta$,
a term $\int \mathrm{d}\mathbf{r} \,
d_{0}(\mathbf{r})|\Delta(\mathbf{r})|^{6}$ with $d_{0}(\mathbf{r})\simeq
-5m^{5}a_{F}^{7}/256\pi$ appears in.  This term contributes a term
$g_{3}|\Psi(\mathbf{r})|^{4}\Psi(\mathbf{r})$ with
$g_{3}=-15\pi^{2}a_{B}^{4}/4m_{B}$ in Eq.~(\ref{eq:GP}); it introduces
the appropriate analogue in Eq.~(\ref{eq:E0}) as well. This three-body
correction to the usual GP equation ($g_{3}$) \cite{PSP03} represents an
effective attraction.  In the composite-boson system, it provides a
first-step correction of the BEC limit en route to the fermionic or BCS
end point.

\begin{figure*}[t]
\centerline{\includegraphics[clip,width=5.in]{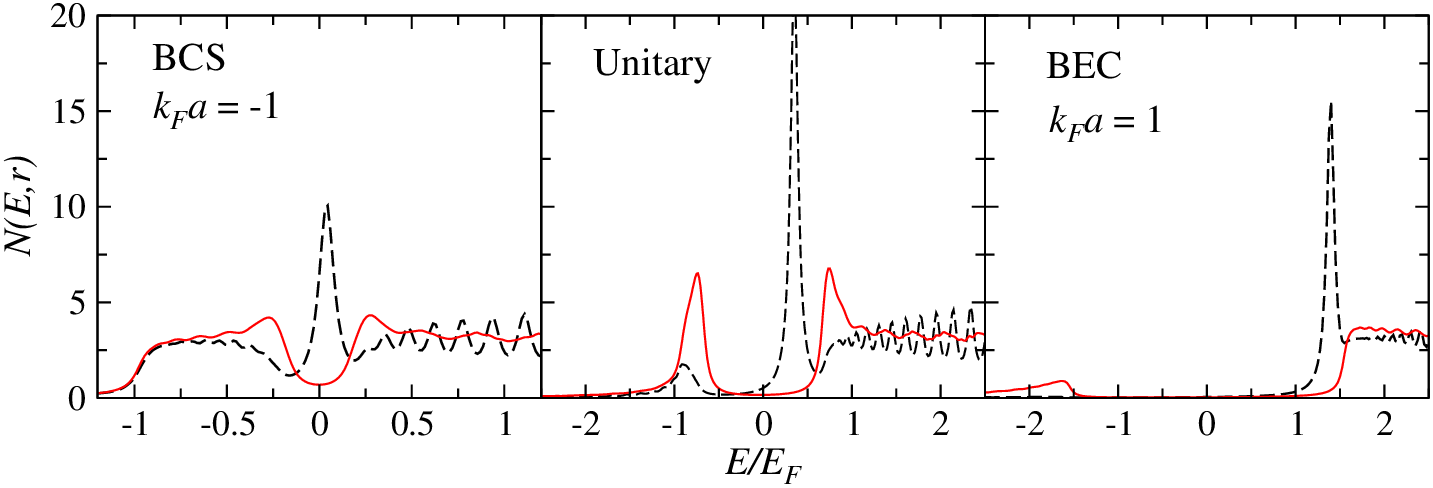}}
\caption{Local fermionic density of states $N(E,r)$ as function of $E$
  for BCS ($k_{F}a{\approx}-1$), unitary and BEC
  ($k_{F}a{\approx}1$) cases, at the center $r=0$ (black dashed
  curves) and radius $r=25/k_F$ (red solid curves) of the vortex core.  The
  bulk value of the gap ${\Delta}_{\infty}$ is 0.21, 0.68, 1.3$E_{F}$,
  respectively.  In the BEC case, $\mu\approx -0.8E_{F}$.  Here $E_F$ is
  the noninteracting Fermi energy at the trap center.}
\label{fig:XEvwg3a}
\end{figure*}

In the presence of a single vortex, the wave function can be written as
$\Psi(\mathbf{r})=f(r)e^{-i\theta}$.  We introduce the BEC correlation
length $\xi_{BEC}$ in the strong-pairing limit as
$(2m_{B}\xi_{BEC}^{2})^{-1}=\mu_{B}=U_{0}f_{0}^{2}$, with
$f_{0}=f(r\rightarrow\infty)$. We rescale $r=\xi_{BEC}\cdot x$ and
$f(r)=f_{0}\cdot\mathrm{X}(x)$ and apply standard boundary conditions
\cite{PS02,PSB03}. 
The results for $\Psi(\mathbf{r})$ in these units are plotted as the
solid lines in Fig.~\ref{fig:XEvwg3}.  As shown in
Fig.~\ref{fig:XEvwg3}a, (and consistent with earlier results in the
literature \cite{PS02,PSB03}), the wavefunction rises smoothly from zero
at the center of the core to its full magnitude at infinity on a length
scale of $\xi_{BEC}$.

An important feature of this figure should be noted.  In this BEC limit
of the BdG equations, the wavefunction is smooth.  This behavior, which
is in contrast to the BCS limit, is a consequence of the absence of
localized fermionic bound states in the core region.  In the weak
coupling limit, as first noted by Caroli et al \cite{CG64}, these bound
states are associated with energy eigenvalues $E_{n}<\Delta_{\infty}$.
Here $\Delta_{\infty}$ is the value of gap function in the bulk, away
from the core. The gap $\Delta(r)$ provides an effective potential well
for the quasiparticles around the vortex core.
%
From Eq.~(\ref{eq:BdG}) we have $E_n \ge -\mu$.
Since $-\mu \gg \Delta_\infty$, these bound states are necessarily
absent in the strong pairing limit .

The first appearance of fermionic properties in our composite-boson
system is via the $g_{3}$-term in the GP equation and the corresponding
term in the zero-temperature energy $E_{0}$.  The effects of this
addition are plotted as dashed lines in Fig.~\ref{fig:XEvwg3} for
(rescaled) $g_{3}^{\prime}=- g_3 f_0^2/U_0 = 0.1$.
This $g_{3}$ contribution represents an
attractive interaction, and, as shown in the figure, leads to a slight
increase in the core size.

One can similarly compute the particle density $n_B(r)$ associated with
composite bosons which is simply related to the wave function as
$n_{B}(r) = |\Psi(\mathbf{r})|^2 = n(r)/{2}$, where $n$ is the
density of fermions.
The density $n(x)$ is plotted in Fig.~\ref{fig:XEvwg3}(b), normalized
at the bulk value $n_\infty \equiv n(\infty)$, as a function of $x$.
As expected the particle density is strictly zero at the core center
in the BEC limit, where there is complete depletion.

The energy cost of a single vortex can also be calculated from
Eq.~(\ref{eq:E0}) and the result is plotted in Fig.~1(c).  The energy
cost per unit length is given by $E_{v}=\frac{\pi
  f_{0}^{2}}{m_{B}}\int_{0}^{R/\xi_{BEC}}\left[
  \left(\frac{d\mathrm{X}}{dx}\right)^{2} + \frac{\mathrm{X}^{2}}{x^{2}}
  +\frac{1}{2}(\mathrm{X}^{2}-1)^{2} \right]xdx$, where $R$ is a cutoff
needed to regularize a calculation of the vortex core energy. In
Fig.~\ref{fig:XEvwg3}(c) the solid line indicates $E_{v}$ as a function
of $R/\xi_{BEC}$.  The shape of the curve at the region $R/\xi_{BEC}>2$
can be fitted to the usual functional form
$E_{v}\propto\ln(D_{G}R/\xi_{BEC})$, where $D_{G}=1.48$.
%
The dashed line in Fig.~\ref{fig:XEvwg3}(c) presents results for $E_{v}$
in the presence of the three body term, where we take
$g_{3}^{\prime}=0.1$.  This correction (red dot-dashed curve) lowers the
vortex energy, as shown in the figure, and it approaches an asymptote as
$R/\xi_{BEC} \rightarrow \infty$.  To understand the details of the core
structure we turn now to numerical solutions of the BdG equation. We
build on our analytical analysis at strong coupling to provide a check
for our numerical algorithms.  Here the physical coupling strength is
controlled by the parameter $1/k_Fa$. We have verified that changes in
our high cutoff energy and associated coupling constant $V$ do not
affect the numerical results.  Our numerical method is very similar to
that in Ref.~\cite{NygaardPRA04}. The chemical potential $\mu$ is
approximated by the homogeneous solution, since the vortex core only
occupies a small portion of the entire system.
Here we begin with a study of the localized (fermionic)
density of states (LDOS), $N(E,r)$, within the core region. There is
considerable interest in the literature in the behavior of the LDOS for
high $T_c$ \cite{TesanovicFranz} as well as low $T_c$ superconductors
\cite{GS91}, since this quantity is accessible through scanning tunneling
microscopy measurements.  $N(E,r)$ is given by $\sum_{n} [u_{n}^2
\delta(E-E_{n})+ v_{n}^2 \delta(E+E_{n})]$.  We ignore, for numerical
simplicity, dependencies of the wave functions on the cylindrical
variable $z$, since these do not lead to qualitative effects.
Integrating $N(E,r)$ over $E \leq 0$ reflects the particle density
distribution $n(r)$ inside the core. Thus, this quantity provides a
means of understanding the density depletion, or lack thereof, inside
the core.  It is essential for arriving at a deeper understanding of the
core region and structure.

In Fig.~\ref{fig:XEvwg3a} we plot $N(E,r)$ inside the core, as a
function of energy $E$ for $r=0 $ and $r=25/k_F $.  The three panels
correspond to BCS (the noninteracting wavevector at the trap center
$k_Fa = -1$), unitary and BEC ($k_Fa = 1$). Rather than showing results
for the strict BCS and BEC regimes, these plots represent a physically
accessible range of magnetic fields.  Because $N(E,r)$ is a
fundamentally fermionic quantity, this information is lost from the
analytical analysis which leads to a Gross-Pitaevskii transcription of
the BEC limit. In the BCS case, the $r=0$ peak at $E=0$ reflects a bound
fermionic state. This state together with the continuum of scattering
states is responsible for the fact that there is no core density
depletion.  For the unitary regime, the lower energy peak in
Fig.~\ref{fig:XEvwg3a} arises from scattering states and appears at
energies near the bulk gap $\Delta_{\infty}$. This quasi-bound state has
energy close to the scattering state continuum and is reflected in the
LDOS by a slight movement of the BCS central peak to the right. The
energy integral of this feature is important in determining the finite
particle density $n(r)$ at the core center.  The peak at positive energy
is a reflection of a quasi-bound state. This unoccupied quasi-bound
state effectively depletes the spectral weight for $E<0$, and
therefore leads to the density depletion within the core.  By the BEC
limit all remnants of fermionic states have disappeared, until one
provides energy large enough to break the pairs.  It can be seen from
the figure that in all three cases at sufficiently large distances from
the core center the fermionic density of states assumes the bulk value.

\begin{figure}
\centerline{\includegraphics[clip,width=3.3in]{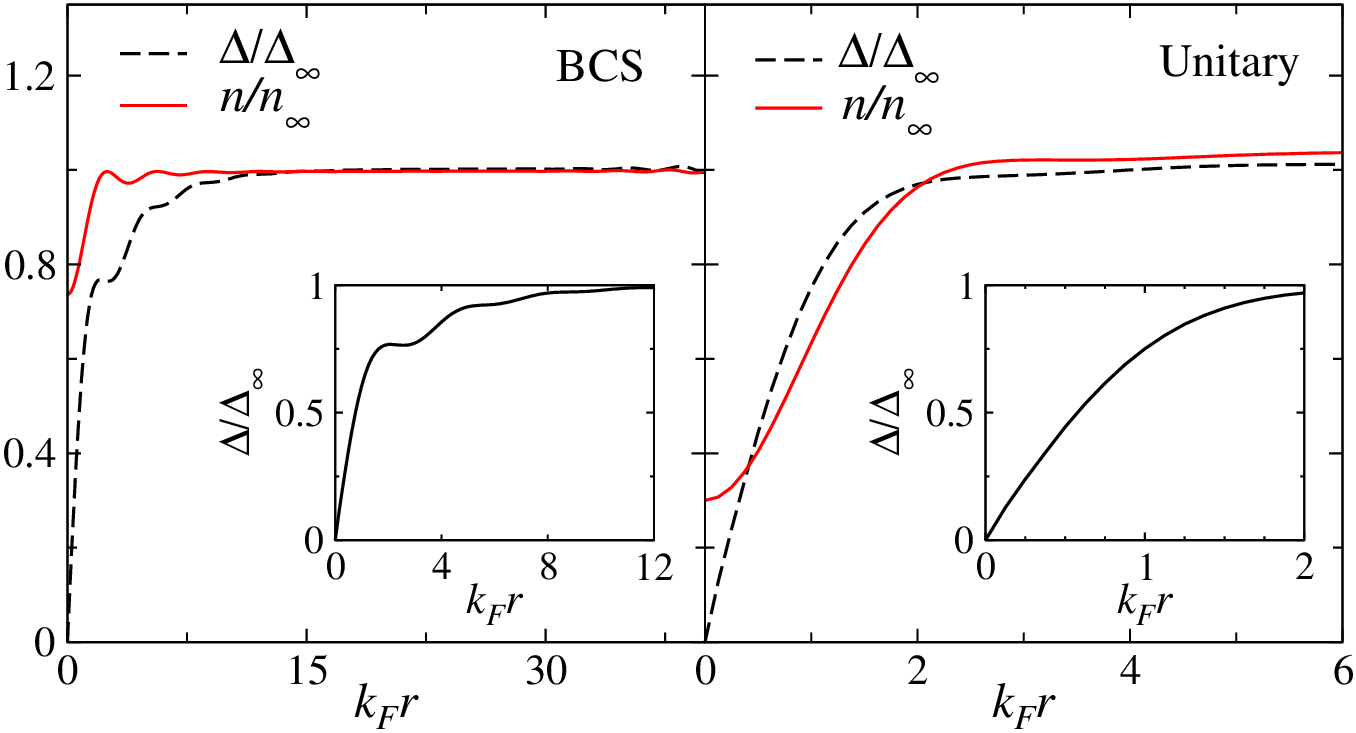}}
\caption{Gap function $\Delta$ (black dashed curves) and particle
  density $n$ (red solid curves) as a function of $r$ for BCS ($k_F a
  = -1$) and unitary cases.  All quantities are normalized by their
  bulk values.  The inserts in the left panels shows detailed behavior
  of the gap near the center of the vortex core. }
\end{figure}

In Fig.~3 we plot the position dependent order parameter $\Delta(r)$
along with the particle density distribution $n(r)$ for the unitary
and BCS ($k_F a = -1$) cases shown in the previous figure.  This
BCS-like case still has a reasonably large bulk gap, so there a
non-negligible depletion at the core. This is in contrast to
arbitrarily weak coupling, where the depletion vanishes.  The small
$r$ oscillations shown here in both $\Delta(r)$ and $n(r)$ reflect the
presence of a true bound state, as in earlier work \cite{NB03}.  The
oscillations at large $r$ are an effect of the finite system size.  

Importantly, in the unitary case, the particle density at the core
center is substantially lower than in the BCS case.  This is a
consequence of the reduced spectral weight seen in the lower peak in the
middle panel of Fig.~2.  Our results are within 20\% of those obtained
in Ref.~\cite{BY03}. In this earlier work a Hartree-Fock correction was
applied to the BdG equations which was argued \cite{Bulgac2} to be the
source of the density depletion.  Here, we interpret this depletion
differently as associated with the behavior of the core excitation
spectra in conjunction with the reduced chemical potential ($\mu <
E_F$).

We thank N. Nygaard for sharing his thesis.  This work
was supported by NSF-MRSEC Grant No.~DMR-0213745.

\textbf{Note} After this work was complete we learned of a related
calculation by Machida and Koyama (Phys. Rev. Lett. 94, 140401 (2005))
which attributed the density depletion within the vortex core at
unitarity to closed-channel bosons.

\bibliographystyle{apsrev}

\end{document}